\documentclass[12pt]{article}
\newcommand{\nn}{\nonumber}
\newcommand{\qq}{\qquad}

\newcommand{\lb}{\label}

\newcommand{\dpsi}{\delta\Psi}
\newcommand{\dphi}{\delta\Phi}

\renewcommand{\P}{\Psi}

\newcommand{\km}{{k-1}}

\newcommand{\tc}{\tilde c}

\newcommand{\ck}{|c^k>}
\newcommand{\ckt}{<c^k|}

\newcommand{\sk}{|s^k>}

\newcommand{\ct}{|\tilde c^k>}
\newcommand{\ctt}{<\tilde c^k|}

\newcommand{\st}{|\tilde s^k>}

\newcommand{\ckm}{|c^\km>}
\newcommand{\ckmt}{<c^\km|}

\newcommand{\skm}{|s^\km>}

\newcommand{\dd}{|d^\km>}

\newcommand{\vd}{\v d}
\newcommand{\vdd}{|\v d^\km>}
\newcommand{\vdt}{<\v d^\km|}

\newcommand{\vb}{\v b}
\newcommand{\vbb}{|\v b^\km>}

\newcommand{\dt}{\tilde d}
\newcommand{\tdd}{|\tilde d^\km>}
\newcommand{\dtt}{<\tilde d^\km|}

\newcommand{\bt}{\tilde b}
\newcommand{\tbb}{|\tilde b^\km>}

\newcommand{\U}{{\hat U}}
\newcommand{\f}{{\rm f}}
\renewcommand{\a}{{\rm a}}
\newcommand{\Uf}{{\hat U}_\f ^k}

\newcommand{\out}{\not\in}

\newcommand{\hf}{{1\over 2}}

\newcommand{\app}{\approx}

\newcommand{\eq}{\equiv}

\newcommand{\be}{\begin{equation}}
\newcommand{\ee}{\end{equation}}
\newcommand{\ba}{\begin{eqnarray}}
\newcommand{\ea}{\end{eqnarray}}

\renewcommand{\d}{\partial}

\newcommand{\de}{\delta}

\newcommand{\ga}{\gamma}

\newcommand{\LL}{{\cal L}}

\newcommand{\OO}{{\cal O}}
\newcommand{\cc}{{\cal C}}

\newcommand{\VV}{{\cal V}}
\newcommand{\WW}{{\cal W}}

\renewcommand{\H}{{\hat H}}
\newcommand{\T}{{\hat T}}

\newcommand{\I}{{ I}}

\newcommand{\w}{\omega}

\renewcommand{\v}{\check}

\newcommand{\pp}{{2m-1}}

\newcommand{\W}{W}
\newcommand{\hh}{{ 1\over\sqrt{h} }}

\newcommand{\F}{\Phi}

\begin{document}
\begin{center}
{\Large \bf An efficient numerical quadrature for the calculation of the potential
energy of wavefunctions expressed in the Daubechies wavelet basis.
}\\
\end{center}
\vspace{1cm}
A. I. Neelov and S. Goedecker \\ \\
{\small 
Institute of Physics, University of Basel, Klingelbergstrasse 82,
CH-4056 Basel, Switzerland \\ \\
E-mail: Alexey.Neelov@unibas.ch, Stefan.Goedecker@unibas.ch \\ \\ \\
{\bf Abstract.\ \ } 

An efficient numerical quadrature is proposed for the approximate calculation
of the potential energy in the context of pseudo potential electronic structure 
calculations with Daubechies wavelet and scaling function basis sets. 
Our quadrature is also applicable in the case of adaptive spatial resolution. 
Our theoretical error estimates are confirmed by numerical test calculations 
of the ground state energy and wave function of the harmonic oscillator in one 
dimension with and without adaptive resolution.
As a byproduct we derive a filter, which, upon application on the 
scaling function coefficients of a smooth function, renders 
the approximate grid values of this function.
This also allows for a fast calculation of the charge
density from the wave function.
}
\section*{1. \quad Introduction.}

Gaussians  and plane  waves are at present the most popular basis  sets for 
density functional electronic structure calculations. Wavelets are a promising 
new basis set that combines most of the theoretical advantages of these two 
basis sets. They can form a systematic 
orthogonal basis set  that allows for adaptivity, the basis functions being
localized both in real (compact support) and in Fourier space.

The first attempts to use 
wavelets in the electronic structure calculations appeared more than 10 years ago.
The first papers we are aware of used the Mexican hat wavelet \cite{Arias0},
\cite{hancho} and the Meyer wavelet \cite{Yamaguchi}. However, these wavelet 
families were soon abandoned because they do not have compact support.
Daubechies \cite{Daub} wavelets were then 
investigated in a series of publications
\cite{weichou}-\cite{fischer}. This basis is orthogonal and has the property of having the highest number of vanishing moments for the given support width, thus combining locality and
approximating power. It is also localized in the momentum space.
To use the Daubechies wavelets in the variational Galerkin method for the Schroedinger equation, 
one has to compute the matrix elements of the kinetic and potential energy operators.
The algorithm for the kinetic part is straightforward \cite{beyl2}. 

The main difficulty is the calculation of 
the potential energy matrix elements \cite{resnik}-\cite{perrier}. They were computed by expanding the potential in terms
of scaling functions, too, and then a convolution was performed with the matrix of products of 
three scaling functions. We will call this "the triple product method": for details,
see the beginning of Section 4. It requires a lot of computer resources; this motivated
alternative approaches. The collocation approach \cite{vasiliev} is not well suited 
for the Daubechies' wavelets since it spoils the favorable convergence rate of variational 
schemes. Another approach \cite{John} involved designing
a quadrature for the product of two scaling functions and a smooth function. 
It decreased the amount of computations in comparison to the triple-product method, 
but not sufficiently.

It follows from the above considerations that the Daubechies basis set can only 
be useful for 
electronic structure calculations if one has a better algorithm for the calculation
of the potential energy than those available at the moment. Such an algorithm will 
be proposed in the present paper.

Due to the above listed problems with the Daubechies family, interest focused 
recently  on the interpolating Deslarier-Dubuc \cite{DD} family 
\cite{Arias1}-\cite{Jac}.
Because the scaling functions of this family are interpolating (cardinal), 
the collocation approximation is much more accurate for them than for the 
Daubechies family. An even more accurate approximation for the potential energy 
is based on a relation with the analytically known overlap matrix elements \cite{Jac}. 

The major disadvantage of the Deslarier-Dubuc wavelets is that they are not orthogonal. 
For very large systems, the dominating term in independent particle electronic 
structure calculations is the orthogonalization of the one particle orbitals.
The prefactor for this dominating cubic term is much smaller if an orthogonal basis 
set is used compared to a non-orthogonal basis set. 
Orthogonal wavelets are in addition also interesting 
candidates for the implementation of linear scaling algorithms \cite{Goed_red}. 

Alpert \cite{Alpert1}, \cite{Alpert2} polynomial multiwavelets overcome the above 
mentioned disadvantages. The potential 
energy can be calculated easily and they are orthogonal. It seems that they 
are the ideal basis set for all electron electronic structure calculations and 
impressive results have been reported \cite{Harr1}-\cite{Harr3}.
The Chui-Lian \cite{chui} family has also been used in the same context \cite{John2}.
Since multiwavelets can represent discontinuous functions, they are 
well suited to represent the electron-nucleus cusp in all electron 
calculations. However, if one uses the Bachelet-Hamann-Schulter \cite{bhs} or Gaussian \cite{gp} pseudopotentials, the wavefunctions and the potential are smooth and this property is not needed. 
Then one may prefer the Daubechies wavelets due to their simplicity and small support length.
For this reason we explore in this paper the use of Daubechies wavelets 
for pseudopotential electronic structure calculations, together with our novel quadrature scheme.
A topic that is closely related to the problem of integrating the potential energy 
is the problem of finding quadrature schemes 
for the product of a scaling function(wavelet) and a smooth function  \cite{John}, \cite{Sweld1}-\cite{john3}. 

The quadrature scheme to be presented in this paper is aimed at applications
in independent particle schemes such as density functional 
theory where one 3-dimensional single particle orbital is provided for each electron in 
the system. For completeness, we will mention that wavelets have also been explored 
as a basis set for high dimensional many-electron wavefunctions \cite{Hack1}-\cite{Hack4}. 
Another approach is to use the so-called Weyl-Heisenberg wavelets that do not have 
multiresolution properties but are related to the structure of the phase space 
\cite{Poir0}-\cite{Poir3}.  Although
these two approaches are promising, any treatment of correlation entails an important 
increase of the numerical effort and such approaches will not allow to treat systems 
with several hundred atoms in the near future.

In the present paper we apply one-scaling-function quadratures to the 
numerical calculation of the potential energy matrix element (\ref{pef}) 
between two smooth functions.  For this purpose 
we developed an algorithm for the reconstruction of grid values of a smooth function
from its scaling function expansion coefficients. This technique might also be used 
in other contexts such as in speech reconstruction. 
Our scheme also provides a way to calculate the density
from the wave function expressed in scaling functions.
The extension of our potential energy quadrature onto the case of adaptive spatial
resolution is then described. 

The paper is organized as follows:

In Section 2 we briefly recall the definition and properties of the Daubechies
wavelet family.

In Section 3 we recall the quadrature of \cite{John},\cite{Sweld1} for the product
of a scaling function and a smooth function.

Using that, in Section 4 we construct the quadrature for the potential energy functional,
i.e., for the product of the form 
\ba
<\,smooth\ function\,|\,potential\,|\,another\ smooth\ function\,>.\lb{pef}
\ea
In subsection 4.1 we derive the quadrature and estimate its error, which
behaves essentially as a square of the error for the wavelet expansion of the smooth functions involved. Then in subsections 4.2-4.3
we prove that the quadrature is exact if the potential and one of the functions in (\ref{pef}) are
polynomials and another function is a scaling function. This confirms
the previous estimation of error and extends it to the case when only one of the functions
in (\ref{pef}) is smooth.

In Section 5 we extend our method onto the case of adaptive resolution, repeating 
the procedure of the previous Section. The errors are again estimated. It is shown that now the main source of error is the boundary between regions with different resolution.

In Section 6 we modify our adaptive quadrature so that in the regions with constant resolution it reduces to the non-adaptive one. This reduces the computational burden.
The price is that now we need to minimize the Rayleigh-Ritz (RR) functional in the space
of smooth functions, i.e., smoothen the cusp of its gradient at the boundary between
regions with different resolution. 

In Section 7 we present a way to compute the density corresponding to a wave function
expressed in scaling functions. 
It uses the approximate wave function values derived by the method described
in Appendix A. This way is fast, but it reproduces multipole moments of the density
only approximately (although with good precision).

Finally, in Section 8 we apply our methods to the calculation of the ground state energy
and wave function of the harmonic oscillator, both with and without adaptivity. 
The results are then compared with those obtained with the RR functional in which
the potential energy was calculated exactly. We consider both the least asymmetric 
and extremal phase Daubechies wavelets and argue that the least asymmetric family is preferable.

\section*{2. \quad The orthogonal wavelets.}

In this work we use the Daubechies \cite{Daub} scaling functions $\phi(x)$ and wavelets $\psi(x)$,
in the dilated and shifted  form:
\ba
\phi_i^k(x)\eq 2^{k/2}\phi(2^kx-i);\qq \psi_i^k(x)=2^{k/2}\psi(2^kx-i).\nn
\ea
where $i,k$ are integer. Sometimes we will also employ the intermediate notation:
\ba
\phi_i(x)\eq \phi(x-i);\qq \phi^k(x)=2^{k/2}\phi(2^kx).\nn
\ea
Our conclusions are the same for the least asymmetric
 and extremal phase \cite{Daub} Daubechies
wavelets, but the least asymmetric ones behave better in the examples we 
considered. Thus we will use the least asymmetric family for the illustration. The graph of the scaling
function of this family of the order 8 is given on Fig. 1.

Since the Daubechies scaling functions and vectors are orthogonal, it is natural to
use the bra and ket notation for them. Then, their orthogonality conditions can
be written as
\ba
<\phi_i^k|\phi_j^k>=\de_{ij};\qq
<\phi_i^k|\psi_j^{k'}>=0,\ \  k'>k;\qq 
<\psi_i^k|\psi_j^{k'}>=\de_{ij}\de_{k\,k'}.\nn
\ea

One can define the following pair of sequences of spaces:
\ba
\VV_k=span\{|\phi_i^k>\};\qq \WW_k=span\{|\psi_i^k>\}\lb{vwk}
\ea
where $\VV_k \ \cap\  \WW_k=\{0\}$.
It turns out that for any $k\geq 0$ the space of square integrable functions $\LL^2$ can be decomposed into the following infinite direct sum:
\ba
\LL^2=\VV_k\oplus \WW_k\oplus \WW_{k+1}\oplus\ldots.\lb{ll2}
\ea

In this paper we will need the refinement relations:
\ba
|\phi_i^{k-1}>=\sum_{j}h_j|\phi_{2i+j}^k>;\qq
|\psi_i^{k-1}>=\sum_{j}g_j|\phi_{2i+j}^k>.\lb{refine}
\ea
They are just a unitary transformation to a new basis; for the spaces (\ref{vwk})
it means that 
\ba
\VV^k=\VV^\km\oplus \WW^\km.\lb{vvw}
\ea
The inverse of (\ref{refine}) is called the forward wavelet transformation,
but we do not need its explicit form in the present text.
 
The decomposition (\ref{ll2}) can be reformulated in the following way: any square integrable function $f(x)$ can be uniquely expanded as
\ba
f(x)=\sum_jc_j^k\phi_j^k(x)+\de f(x);\qq
\de f(x)=\sum_{q=k}^\infty\sum_jd_j^{q}\psi_j^{q}(x).\lb{fw}
\ea

The Daubechies-$2m$ scaling functions are a basis of degree $m-1$. This means that every
polynomial of degree less than $m$ is contained in $\VV_k$. 
Therefore \cite{Unser1}, the tail part of the series (\ref{fw}) behaves as
\ba
\de f(x)=\OO(h^{m});\qq h=2^{-k}\lb{defo}
\ea
with respect to the $\LL^2$ norm. The $\OO$ notation corresponds to the limit of $k\to\infty$. For the polynomials of degree less than $m$, $\de f(x)=0$.

The methods to be presented can be extended to other wavelet families, not 
necessarily orthogonal, but in our opinion, the Daubechies family is the optimal choice
for the electronic structure calculations.

\section*{3. \quad Quadrature for orthogonal wavelets.}

A wavelet quadrature is determined by the set of coefficients $w_l$ such that
for a smooth function $G(x)$, the integral $\int G(x) \phi_r^{k}(x)dx$ is
approximated by $\sqrt{h}\sum_lw_lG_{l+r}^k$, where $G_j^k\eq G(2^{-k}x)$. 
The $\sqrt{h}$ factor comes from the normalization of the scaling function.
The degree of accuracy of a quadrature formula is
$M$ if it yields the exact result for every polynomial of degree less than or equal to $M$.
This is equivalent to the condition \cite{John}
\ba
\sum_l \w_l l^s=M_s;\qq M_s\eq\int y^s\phi(y) dy\lb{Ms}
\ea
where $M_s$ are the scaling function moments.

If the quadrature filter is of degree $M$ and $G(x)$ belongs to $\cc^{M+1}$ then \cite{John},\cite{Sweld1}
\ba
\int G(x) \phi_r^{k}(x)dx&=&\sqrt{h}\sum_lw_lG_{l+r}^k+\OO(h^{M+3/2})=
\sqrt{h}\sum_s w_{s-r} G_s^k+\nn\\
&+&\OO(h^{M+3/2});\qq 
h=2^{-k}.\lb{fphi}
\ea
In this paper we will usually work with the uniform quadrature (of degree $M$) set forth in \cite{John}:
\ba
w_l=\sum_{r=0}^{M}P_{lr}M_r.\lb{wl}
\ea

It will be supposed that $M=2m-1$ for the Daubechies-$2m$ wavelets, and $l=1-m..m$.
In (\ref{wl}), the Lagrange polynomials of degree $M$ are used:
\ba
P_l(y)={\prod_{j=1-m}^{m}}'{y-j\over l-j}\eq \sum_{r=0}^{2m-1}P_{lr}y^r;\qq
P_{lr}={1\over r!}{d^rP_l(y)\over d y^r}\bigg|_{y=0}.\nn
\ea
The nonzero values of the filters $w_l$ for $m=3..6$ are shown at the Table 1.

\begin{table} 

Table 1. The values of the filters $w_l$ for the Daubechies-$2m$ least asymmetric filters; $m=3..6$.\\
\begin{footnotesize}
\begin{tabular}{ccccc}

\hline \hline l  & {\hskip 0.2 cm} Daubechies-6 {\hskip
0.2 cm} & {\hskip 0.2 cm} Daubechies-8 {\hskip 0.2 cm} &{\hskip 0.2 cm} Daubechies-10 \ {\hskip 0.2 cm} & {\hskip 0.2cm} Daubechies-12 {\hskip 0.2 cm}
\\
\hline  \\ 
-5 &                     &                     &                     &  0.0000754232174770\\
-4 &                     &                     &  0.0003712028220936 & -0.0011760498174610\\
-3 &                     &  0.0026299127476935 & -0.0046529756260417 &  0.0104347966396891\\
-2 &  0.0858797754503928 & -0.0377927339236569 &  0.0306436002784248 & -0.0340901829704789\\
-1 &  1.0472376804223309 &  0.0755988357512099 & -0.1207447752890374 & -0.0067678682684262\\
 0 & -0.1886782932535312 &  0.9999560903030736 &  0.1338108260452157 &  1.0005931732054807\\
 1 &  0.0795781221430145 & -0.0794124676160406 &  0.9123169219278740 &  0.0041859363010669\\
 2 & -0.0288721312776034 &  0.0451427040622791 &  0.0109419516584456 &  0.0351468153360141\\
 3 &  0.0048548465153963 & -0.0069875964135745 &  0.0393078583967683 & -0.0096794739531791\\
 4 &                     &  0.0008652550890159 & -0.0022599250999316 &  0.0015648660417616\\
 5 &                     &                     &  0.0002653148861886 & -0.0003139771845937\\
 6 &                     &                     &                     &  0.0000265414526497\\
\\
\hline \hline

\end{tabular}
\end{footnotesize}
\end{table}

Fig. 1 also contains the values of $w_l$ for $m=4$ and $M=7$, compared with the graph
of the corresponding scaling function (least asymmetric Daubechies-8). One sees that the filter values are close to the scaling function values at integer points.

\section*{4. \quad The potential energy functional for orthogonal wavelets.}

An important step in the electronic structure calculation is to find the eigenspectrum
of a one-particle Schroedinger equation. Although our approach can be extended to the three-dimensional systems, in this paper we will consider only one-dimensional
ones:
\ba
\H\Psi(x)=E\Psi(x);
\qq \H=-\hf{\d^2\over\d x^2}+ V(x)\eq \T+\U;
\qq \int d x \P^2(x)=1.\lb{hpsi}
\ea
This equation can be solved by representing a trial wavefunction as a linear combination
of finite elements \cite{Strang}. In the case considered in this paper, these
finite elements are the Daubechies scaling functions (wavelets). Thus we select the trial function in the form
\ba
\P_I(x)=\sum_i c_i^k\phi_i^k(x)\lb{pi1}
\ea
 which can also be represented by the ket vector $\ck=\sum_{i=-N}^N c_i|\phi_i^k>$
with $N=L/h=2^{-k}L$. In this paper we use nonperiodic boundary conditions, although we could  use the periodic ones too. Then if the Rayleigh-Ritz functional is defined as
\ba
R(\{c_i^k\})&=&{\ckt \H\ck\over \ckt c^k>}= {T(\{c_i^k\})+U(\{c_i^k\})\over \ckt c^k>};\lb{RR}\\
T(\{c_i^k\})&\eq& \ckt \T \ck;\qq
U(\{c_i^k\})\eq \ckt \U\ck\lb{u}
\ea
the variational ground state energy and wave function expansion coefficients have the form
\ba
E_0=\min R(\{c_i^k\});\qq |c^k_0>=\arg\min R(\{c_i^k\}).\lb{min}
\ea
The other wavefunctions are obtained in the same way using the Lagrange multipliers.

Since the Daubechies-$2m$ scaling functions have degree $m-1$ (see Section 2), the ground state energy
error behaves \cite{Strang} as
\ba
\de E\eq E_0-E_{ex}=\OO(h^{2m-2})\lb{eex}
\ea
where $E_{ex}$ is the exact ground state energy of (\ref{hpsi}).

The kinetic energy has the form \cite{Gbook}
\ba
T(\{c_i^k\})=-{1\over 2h^2}\sum_{i,j} a_{i-j}c_i^k c_j^k\nn
\ea
where $a_l$ is the kinetic energy filter. 

The potential energy (\ref{u}) can be written in the conventional form as
\ba
U=\int \P_I(x)V(x)\P_I(x)dx.\nn
\ea
For the calculation of the excited states we have to evaluate the more  general potential energy bilinear form:
\ba
U=\int \F_I(x)V(x)\P_I(x)dx\lb{up}
\ea
where the wave function $\F_I(x)$ has the same form (\ref{pi1}):
\ba
\Phi_I(x)=\sum_i c_i^k\phi_i^k(x)  .\lb{pi2}
\ea 

In the electronic structure calculations the exact analytical form of the potential $V(x)$
is not known. Any approximation of the potential should 
invoke energy errors that alter the Rayleigh-Ritz functional (\ref{RR}) as little as
possible; otherwise the minimization with an approximate functional will not converge
to the true minimum. We choose the following quantitative criterion for this:
the error in the potential that arises from approximation should be small 
compared to (\ref{eex}). When the grid parameter $h$ is small, there is a convenient way to ensure that: we require that the approximation error behaves as $h^{2m}$. Note that, e.g.,
the collocation approximation \cite{vasiliev} does not fulfill the latter requirement.

The most natural way to approximate the potential is to expand it in the interpolating scaling functions  
$\phi^{kI}_i(x)$ \cite{DD},\cite{Gbook} of degree $2m$:
\ba
V(x)&=&\sum_i V_i^k\phi^{kI}_i(x)+\de V(x);\qq \de V(x)=\OO(h^{2m});\lb{vint}\\
\phi^{kI}_i(x)&\eq& \phi^\I(x/h-i);\qq
V_i^k\eq V(ih);\qq h=2^{-k}.\nn
\ea
where the error estimate (\ref{vint}) holds for sufficiently smooth potentials.
Substituting the expansions (\ref{pi1}),(\ref{pi2}),(\ref{vint})  into (\ref{up}), we get:
\ba
U&=&\sum_{ijl}c_i^kV_j^ks_l^k \int \phi_i^k(x)\phi_j^{kI}(x)\phi_l^k(x)dx+\OO(h^{2m})=
\sum_{ijl}c_i^kV_j^ks_l^kI_{i-j,l-j}+\lb{ui}\\
&+&\OO(h^{2m});\qq
I_{rs}\eq\int \phi_r(x)\phi^\I(x)\phi_s(x)dx\lb{irs}
\ea
where (\ref{irs}) is the triple product matrix mentioned in the Introduction.
Unfortunately, the formula (\ref{ui}) requires too many floating point operations (flops) per grid point. In the simplest case when $s_i^k$ coincide with $c_i^k$, the leading term in that number is $3/2 N^2$ in one dimension and $2N^4$ in three dimensions. $N$ is the dimension of the matrix (\ref{irs}), and eigenvalue decomposition of the matrix (\ref{irs}) is used in the 3D case. For the Daubehies-$2m$ scaling functions, $N=2m-2$, so the leading term in $m$ of the number of flops per gridpoint is $6m^2$ in 1 dimension and $32 m^4$ in 3 dimensions. This is unacceptable for the realistic values of $m$ of the order of $10$.
 
 Therefore we need some approximation that would decrease
the amount of computations, but still invoke the error that asymptotically behaves as $\OO(h^{2m})$, at most.
Such an approximation will be described in the next subsection.

\subsection*{\it 4.1 \quad The quadrature for the case of smooth wave functions.}

Let us consider two smooth functions $\F(x)$, $\P(x)$ together with their
approximate wavelet expansion $\F_I(x)$, $\P_I(x)$ (\ref{pi1}),(\ref{pi2}). The expansion 
coefficients $c_i^k$ and $s_i^k$ are obviously given by
\ba
c_i^k=\int \phi_i^k(x)\Phi(x) dx;\qq
s_i^k=\int \phi_i^k(x)\Psi(x) dx .\lb{cik}
\ea
Then, according to (\ref{fw}),(\ref{defo}),
\ba
\F(x)&=&\F_I(x)+\dphi(x) ;\qq
\P(x)=\P_I(x)+\dpsi(x)
 ;\lb{depsi1}\\
 \dphi(x)&=&\sum_{k'> k}\sum_i d_i^{k'}\psi_i^{k'}(x)=\OO(h^{m});\ \ \
\dpsi(x)=\sum_{k'> k}\sum_i b_i^{k'}\psi_i^{k'}(x)=\OO(h^{m}).\lb{depsi2}
\ea
From (\ref{depsi1}) we get 
$$
 \F_I(x)=\F(x)-\de\F(x) ;\qq \P_I(x)=\P(x)-\de\P(x).
$$ 
Plugging this expression into (\ref{up}),
 we can get after some easy calculations:
\ba
U&=&U_A+\de U_1+\de U_2;\qq U_A\eq\int \F_I(x)V(x)\P(x)dx;\lb{udeu}\\
\de U_1&\eq& \int \dphi(x)V(x)\dpsi(x)dx;\qq
\de U_2 \eq-\int \F(x)V(x)\dpsi(x)dx
 .\lb{up3}
\ea
In the next Section we will prove that
\ba
\de U_1+\de U_2=\OO(h^{2m})\lb{deu}
\ea
from the local properties of scaling functions. In this Section we will show 
for the two error terms separately that 
\ba
\de U_1=\OO(h^{2m});\qq \de U_2=\OO(h^{2m}).\lb{dvd}
\ea
The first estimate of (\ref{dvd}) follows from (\ref{depsi2}) and the fact that individual wavelet coefficients $d_i^{k'}$, $b_i^{k'}$ from the
tail parts behave as $2^{-k'(m+1/2)}$ \cite{Sweld1}. If the potential $V(x)$ is bounded
 then it follows from the Schwarz inequality:
\ba
\biggl|\int \dphi(x)V(x)\dpsi(x)dx\biggr|\leq max|V|\int |\dphi(x)||\dpsi(x)|dx
\leq max|V|\|\dphi\| \|\dpsi\|.\nn
\ea

To prove the second part of (\ref{dvd}), let us define the function $H(x)=\Phi(x)V(x)$. It can also be interpolated by scaling functions:
\ba
H(x)&=&H_I(x)+\de H(x);\qq
H_I(x)=\sum_i h_i^k\phi_i^k(x);\nn\\
h_i^k&=&\int \phi_i^k(x)H(x) dx;\qq
\de H(x)=\sum_{k'> k}\sum_i q_i^{k'}\psi_i^{k'}(x)=\OO(h^{m})
.\nn
\ea
Then, $\int H_I(x)\de \P(x)dx=0$, and
\ba
\de U_2=-\int \de H(x)\dpsi(x)dx=\OO(h^{2m}).\nn
\ea

We are left with the approximation (\ref{udeu}). Substituting (\ref{pi2}) into it, we get
\ba
U_A=\sum_i c_i^{k}\int \phi_i^{k}(x)G(x)dx\lb{phig}
\ea
where $G(x)\eq V(x)\Psi(x)$. The scalar products in (\ref{phig})
can now be determined by applying the scaling function quadrature (\ref{fphi}):
\ba
U_A[\P_I]&=&\sum_i c_i^{k}\biggl[
\sqrt{h}\sum_s w_{s-i} G_s^k+\OO(h^{2m+1/2})
\biggr]=\sum_sp_sG_s^k+\OO(h^{2m})\lb{uapi}
\ea
where 
\ba
G_q^k=G(hq)=V(hq)\P(hq)=V_q^k\P_q^k\nn
\ea
are the grid values of the function $G(x)$. As seen from (\ref{uapi}), the quantity
\ba
p_s^k&\eq&\sqrt{h}\sum_i c_i^k w_{s-i}\nn
\ea
plays the role of the quadrature for the whole function $\F_I$.

Now we need to find the grid values of the smooth function $G(x)$. 
We suppose that we know the grid values $V_q^k$ of the smooth potential. 
Then it remains to
reconstruct the grid values $\P_i^{k}$ of the unknown wave function from
the known coefficients $s_i^{k}$.

The easiest way to obtain the grid values would be by using (\ref{pi1}):
\ba
\P(jh)\app\P_I(jh)=\sum_i s_i^{k}\phi_i^{k}(jh).\lb{pq}
\ea
Since the Daubechies scaling functions are not very regular, the value 
of a scaling function at a real space grid point does not very well 
represent the behavior of the scaling function in a small interval around 
this grid point. As a consequence of this and 
in accordance with (\ref{defo}), the error in (\ref{pq}) behaves
like $\OO(h^{m})$. 
which is much worse than (\ref{deu}) and we therefore discard this possibility. 
We will instead introduce some smoothed grid values $\bar\P_q^{k}$ which better 
represent the average behavior of $\P_I$.  In appendix A it will be shown that there 
exists a filter $W_j$ such that the smoothed grid values can be obtained by a convolution 
from the scaling function expansion coefficients
\ba
 \bar\P_q^{k} \eq  {1\over\sqrt{h}}\sum_t W_{q-t}s^{k}_t=
{1\over\sqrt{h}}\sum_j W_js^{k}_{q-j} \lb{vpq}
\ea
and that the error behaves as
\ba
\P_q^{k} = \bar\P_q^{k}+\OO(h^{2m}).\nn
\ea
Substituting everything back into (\ref{uapi}),we get
\ba
U_A&=&
\sum_{s} p^k_sV^k_s\bar\P^k_s+\OO(h^{2m}).\lb{pvpsi}
\ea
Taking together (\ref{udeu}), (\ref{deu}) and (\ref{pvpsi}), one gets:
\ba
U=U_\f +\OO(h^{2m});\qquad U_\f \eq \sum_{s} p^k_sV^k_s\bar\P^k_s\lb{uuf}
\ea
where "$f$" stands for "filters".

As explained in Appendix A, in the case of the Daubechies scaling functions the quadrature filter
$w_l$ and the reconstruction filter $W_l$ are identical. Then (\ref{uuf}) assumes the 
following simple form:
\ba
U_\f \eq h\sum_{s}\bar\F^k_s V^k_s\bar\P^k_s;\qq 
\bar\F_q^{k}&\eq& {1\over\sqrt{h}}\sum_t w_{q-t}c^{k}_t.\lb{ufd}
\ea 
This formula can be computed very fast: when $c_i^k=s_i^k$, the number of flops per grid point is just $4m+3$. One can use it in three dimensions too.
In that case the quadrature and reconstruction filters are tensor products of the one-dimensional ones. Thus one only needs three convolutions with filters of length $2m$ per grid point for the calculation of the potential energy, thus $12m+3$ flops. This is clearly better than the $32m^4+\ldots$ result for the triple product method.

The formulas (\ref{uuf}),(\ref{ufd}) have a global nature; they characterize the approximation error over the whole domain of the wave function.  In the next two subsections we will choose a local point of view instead, and prove that Eqs. (\ref{uuf}),(\ref{ufd}) hold for unbounded $V(x)$ too.

\subsection*{\it 4.2 \quad The matrix elements of the potential energy in the case of the Daubechies family.}

If we substitute the wavefunction expansion (\ref{pi1}),(\ref{pi2}) into the energy expression (\ref{up}), we get
\ba
U=\sum_{i,j}c_i^k s_j^kU_{ij};\qquad U_{ij}\eq\int dx \phi^k_i(x)V(x)\phi^k_j(x).\lb{uij}
\ea

Similarly, the approximate energy (\ref{ufd}) in the case of the Daubechies family has the form
\ba
U_\f =\sum_{i,j}c_i^k s_j^kU_{ij}^\f   ;\qquad U_{ij}^\f   \eq \sum_l V_l^kw_{l-i}w_{l-j}\lb{ua2}
\ea
where $w_l$ is the quadrature filter for the Daubechies family.

It is enough to consider the matrix elements $U_{0q}$; $U^\f_{0q}$ since others
can be obtained by shifting the potential. Let us assume that $V(x)=x^t$. Then,
after going to the variable $y=x/h$ under the integral one obtains:
\ba
U_{0q}=\int \phi^k(x)x^t\phi_q^k(x)dx=h^tK_{qt};\qq
K_{qt}\eq\int dy\phi(y)\phi_q(y)y^t.\lb{kqp}
\ea
Similarly, one can check that
\ba
U^\f_{0q}=h^tK_{qt}^\f;\qq
K_{qt}^\f \eq\sum_l w_{l-q}w_l\, l^t.\lb{ufk}
\ea
Now, if the potential $V(x)$ is a smooth function, one can expand it
into a Taylor series around the origin:
\ba
V(x)=\sum_{p=0}^\pp{V^{(p)}(0)\over p!}x^p+\OO(x^{2m}).\lb{vt}
\ea

Our aim is to use wavelets for the electronic structure calculations with
pseudopotentials, and the local part of the Gaussian pseudopotentials introduced in \cite{gp} is smooth (has infinite number of continuous derivatives). The smoothness of the potential is best exploited if the grid is fine enough. If necessary, one can also increase the accuracy of our approximation without changing the number of basis functions, by going to the doubly dense real space grid. This is discussed in more detail in the very end of subsection 5.1.
The multiscale nature of wavelets allows one to adapt the grid resolution locally, which will be described in section 5. 

Note however that one needs only the grid values of the potential for the actual calculations
(according to Eq. (\ref{ufd})). The Taylor expansion of the potential (\ref{vt}) is presented
here only to analyze the errors.

Plugging the expansion (\ref{vt}) into (\ref{uij}) and
taking into account (\ref{kqp}), we get:
\ba
U_{0q}=\sum_{p=0}^\pp{V^{(p)}(0)\over p!}h^pK_{qp}+\OO(h^{2m}) .\lb{uk}
\ea
In the same way, plugging (\ref{vt}) into the right equation of (\ref{ua2}) and
using (\ref{ufk}), one obtains
\ba
U_{0q}^\f   =\sum_{p=0}^\pp{V^{(p)}(0)\over p!}h^pK_{qp}^\f +\OO(h^{2m}). \nn
\ea

It is in the coefficients $K_{qp}$ that the approximation differs from the exact calculation. 
Eq. (\ref{ufk}) can be considered as a quadrature approximation of the integral (\ref{kqp}).

It would be convenient if the values of $K_{qp}$ and $K_{qp}^\f $ coincided. It would explain
the smallness of error in (\ref{pvpsi}), since in that case, $U_{ij}$ and $U_{ij}^\f   $ would
differ only by $\OO(h^{2m})$. However, for the cases
we checked (Daubechies-6 to 18) the values of $K$ turned out to be different. Thus, the smoothness of the potential alone
does not explain the smallness of error (\ref{uuf}). In the next subsection we will see that 
one needs also the smoothness of the wavefunction to explain that.

Note that the strategy used in \cite{John} was to find an approximation of the matrix (\ref{uij}) such that the corresponding
$K_q^p$ be exactly equal to (\ref{kqp}), i.e., that the approximation is
exact for the polynomial potentials. This is a flexible scheme but it requires more computer resources than 
ours. The reason is that in the scheme of \cite{John} one has to apply two-index 
filters at each point.
Therefore we have the same problem as with the triple product formula (\ref{ui}): the number
of flops per grid point is
$~N^2$ in one dimension and $~N^4$ in three dimensions, where $N$ is the dimension of
the matrix which plays the role of (\ref{irs}).

\subsection*{\it 4.3 \quad The gradient of the potential energy in the case of the Daubechies family.}

The gradients of the exact and approximate energies (\ref{uij}) and (\ref{ua2}) w.r.t. $c_i^k$ are
\ba
{\d U\over\d c_i^k}=\sum_{j}s_j^kU_{ij};\qq
{\d U_\f \over\d c_i^k}=\sum_{j}s_j^kU_{ij}^\f   \lb{grad} 
\ea
where the matrix elements are discussed in the previous subsection. In the case of the
potential energy functional, with the coefficients $c_i^k$ instead of $s_i^k$ in 
(\ref{uij}) and (\ref{ua2}), there should be an additional factor of 2 in (\ref{grad}).

From (\ref{uuf}) one could guess that 
\ba
{\d U\over\d c_i^k}={\d U_\f \over\d c_i^k}+\OO(h^{2m+1/2})\lb{ggf}
\ea
since there are no preferred points in space for the energy expressions (\ref{uij}) and (\ref{ufd}),
and thus the energy error should be "smeared" smoothly over the grid points. 
Later in this
subsection we will see that (\ref{ggf}) is indeed satisfied.

The condition (\ref{ggf}) is very important. It means that if we minimize $U_\f $ using some numerical method
(steepest descent etc.) then the gradients along which we change the wavefunctions will be 
very close for the exact and approximated energies, and thus the results of minimization,
the two ground state wavefunctions, will also be close.

On the other hand, Eq. (\ref{uuf}) that was derived above only for a bounded $V(x)$, follows from (\ref{ggf}). What is more, it follows from (\ref{ggf}) that
(\ref{uuf}) is satisfied even if we do not require the function $\F(x)$ to be smooth.

At first we will prove (\ref{ggf})
for the partial case of $i=0$, $\P(x)=x^l$ and $V(x)=x^t$, from which we then will easily
make extension onto the general case. With the above assumptions, acting similarly to the proof of (\ref{csm}) one can check that
\ba
s_j^k=h^{l+1/2}\sum_{u=0}^lC_l^uj^uM_{l-u}\lb{sjk}
\ea
where $C_l^u$ are the binomial coefficients, and $M_s$ are the scaling
function moments (\ref{Ms}).

Substituting (\ref{kqp}) and (\ref{sjk}) into (\ref{grad}), one obtains:
\ba
{\d U\over\d c_0^k}=h^{l+t+1/2}A_{lt};\qq
A_{lt}\eq \sum_{u=0}^lC_l^uM_{l-u}\sum_j K_{jt}j^u.\lb{a}
\ea

Similarly, in the case of the approximate energy,
\ba
{\d U_\f\over\d c_0^k}=h^{l+t+1/2}A_{lt}^\f;\qq
A_{lt}^\f\eq \sum_{u=0}^lC_l^uM_{l-u}\sum_j K_{jt}^\f j^u.\lb{af}
\ea
 
The coefficients (\ref{a}), (\ref{af}) satisfy the following equalities:
\ba
A_{lt}&=&M_{l+t},\qq l,t<m;\qq A_{l0}=M_l,\qq l<2m;\lb{am}\\
A_{lt}^\f&=&M_{l+t},\qq t+m<2m\lb{afm}
\ea
where $M_s$ are the scaling function moments (\ref{Ms}).
Their analytical proof can be found in Appendix B.

Since we were unable to find an analytical proof of (\ref{am}) for 
the remaining relevant values of indices:
\ba
A_{lt}&=&M_{l+t},\qq m<t<2m\lb{amm}
\ea
we checked it numerically with Mathematica for the Daubechies wavelets of the orders from 6 to 18. We checked both the extremal phase and least asymmetric Daubechies families.
The relative error of (\ref{amm}) can be made arbitrarily small by increasing the precision
of wavelet filters. We went down to $10^{-40}$.

Thus we see that for the Daubechies wavelets,
\ba
A_{lt}=A_{lt}^\f=M_{l+t};\qq
{\d U\over\d c_0^k}={\d U_\f\over\d c_0^k}=h^{l+t+1/2}M_{l+t};\qq l+t<2m\lb{uu}
\ea
for arbitrary polynomial $V(x)$ and $\Psi(x)$ with 
the sum of their degrees less than $2m$. The second equality follows from (\ref{a}),(\ref{af}).

In the remaining part of this subsection we will infer (\ref{ggf}) from (\ref{uu}).
Let us consider the exact gradient. It is given by the left formula (\ref{grad}). The matrix element $U_{0j}$ 
has been calculated in the previous subsection. Now let us consider the scaling function expansion coefficient $s_j^k$ (\ref{cik}) for a smooth function $\P(x)$. Since the latter is smooth, it can be expanded into the Taylor series at the origin:
\ba
\P(x)=\sum_{p=0}^\pp{\P^{(p)}(0)\over p!}x^p+\OO(x^{2m}).\lb{pt}
\ea
Plugging (\ref{pt}) into (\ref{cik}) and using (\ref{sjk}), we get:
\ba
s_j^k=\sum_{p=0}^\pp{\P^{(p)}(ih)\over p!}h^{p+1/2}\sum_{t=0}^p C_p^tq^{p-t}M_t
+\OO(h^{2m+1/2}).\lb{cjk2}
\ea
Substituting (\ref{cjk2}) and (\ref{uk}) into (\ref{grad}), we get:
\ba
{\d U\over\d c_0^k}
&=&\sum_{s=0}^\pp {h^{s+1/2}\over s!}
\sum_{p=0}^sC^p_s\P^{(p)}(0)V^{(s-p)}(0)A_{p,s-p}+\OO(h^{2m+1/2})=\nn\\
&=&\sum_{s=0}^\pp {h^{s+1/2}\over s!}G^{(s)}(0)M_s+\OO(h^{2m+1/2})=
g_0^k+\OO(h^{2m+1/2})\lb{gaa}
\ea
where we have used (\ref{uu}). As before, $G(x)\eq V(x)\Psi(x)$, and
$g_i^k\eq \int G(x)\phi_i^k(x) dx$.
The analogue of (\ref{gaa}) 
for an arbitrary gradient component ${\d U\over\d c_i^k}$ can be obtained by
replacing all the 0 by $ih$. Then our error estimate for the potential energy of Eq. (\ref{deu}) follows from the formula
\ba
U=\sum_i c_i^k{\d U\over\d c_i^k}.\lb{ugrad}
\ea

For the gradient of the approximate energy one can exactly repeat the above steps,
just using $K_{jt}^\f$ instead of $K_{jt}$ (since one starts from (\ref{ufk}) instead of (\ref{uk})).
The result will be identical to (\ref{gaa}); thus, (\ref{ggf}) is proved.

\section*{5. \quad The case of adaptive resolution.}

In the actual calculations, most of the wave function coefficients for wavelets on fine levels are very small. An advantage of wavelets is that the coefficients smaller than certain threshold can be set to zero \cite{John0}, \cite{tymczak}; this is called dynamical adaptivity. We will use here the so called static adaptivity when one constrains
the fine wavelets to be zero at the some parts of configuration space - e.g., far from the
atomic cores. This effectively leads to grid resolution slowly varying in real
space. Since our quadrature approximation involves some extra care of the boundaries
between regions of different grid resolution (see below), it is more suited for the
static adaptivity where one can make sure that these boundaries have simple form.

The simplest two-level example will be considered below.

\subsection*{\it 5.1\quad The energy error for the adaptive approximation.}

In this Section we will extensively use the bra and ket notation for the 
wavefunctions (\ref{pi1}),(\ref{pi2}):
\ba
\ck\eq \sum_i c_i^k|\phi_i^k>;\qq  \sk\eq \sum_i s_i^k|\phi_i^k>   .\lb{ck}
\ea
The ket vectors belong to the space $\VV_k$ (\ref{vwk}). It is finite-dimensional after we apply boundary conditions,
whether periodic or non-periodic.

The energy expressions (\ref{uij}), (\ref{ua2}) then can be written as
\ba
U=\ckt \U\sk;\qq U_\f =\ckt \Uf\sk.\lb{uvec}
\ea
Note the difference between the exact and approximate operators. $\U$ is 
the operator of multiplication by $V(x)$ in a Hilbert space. On the other hand,
$\Uf$ is an operator in the finite-dimensional space $V_k$, taking into
account the boundary conditions. Its matrix elements are given by (\ref{ua2}).

In the present Section we will consider two resolution levels simultaneously.
In addition to the resolution level $k$, considered up to now, we have 
a more coarse level $k-1$ together with the vectors associated with it:
\ba
\ckm\eq \sum_i c_i^\km|\phi_i^\km>;\qq  \dd\eq \sum_i d_i^\km|\psi_i^\km>   .\lb{css}
\ea
For some time we will discuss only the vectors corresponding to $\Phi_I(x)$; those
related to $\Psi_I(x)$ satisfy the same relations with the symbols $c,d$ replaced
by $s,b$ where appropriate.

Similarly to (\ref{cik}), the coefficients in (\ref{ck}), (\ref{css}) can be expressed in terms of the smooth function $\F(x)$:
\ba
c_i^k&=&\int \phi_i^k(x)\Phi(x) dx;\qq
c_i^{k-1}=\int \phi_i^{k-1}(x)\Phi(x) dx;\nn\\
d_i^{k-1}&=&\int \psi_i^{k-1}(x)\Phi(x) dx  .\nn		       
\ea

The bra and ket notation is convenient because in accordance with (\ref{vvw}), the forward and backward wavelet transformations can be written as
\ba
\ck=\ckm+\dd.\lb{fbt}
\ea
The left and right parts of the above equality differ only in the choice of basis
functions: $\{\phi_i^k\}$ at the left and $\{\phi_i^\km,\psi_i^\km\}$ at the right.

The adaptive approximation consists of replacing 
the second relation in (\ref{css}) by
\ba
\tdd\eq \sum_{i\in D} d_i^\km|\psi_i^\km>. \lb{ddt}
\ea
In this paper we 
consider the so-called static adaptivity, where one chooses some fixed predetermined 
region $D$ of space
such that outside $D$ the wavelet coefficients $d_i^{k}$ are known to be small and can therefore be neglected. We will call $D$ the fine region.

Replacing $\dd$ in (\ref{fbt}) by (\ref{ddt}), one can form the adaptive vector
\ba
 \ct\eq\ckm+\tdd .\lb{ct1}
\ea
In other words, if we forward transform $\ck$, then discard the wavelet part 
outside the domain $D$ according to Eq. (\ref{ddt}) and 
then backward transform the result, then we get $\ct$.

The discarded wavelet part can also be written as a vector:
\ba
\vdd\eq \sum_{i\out D} d_i^\km|\psi_i^\km>\lb{ddisc}
\ea
so that, taking into account (\ref{ddt}),
\ba
\dd=\tdd+\vdd.\lb{ddd}
\ea
The exact energy expression (\ref{uvec}) in the adaptive case assumes the form
\ba
U^\a  =\ctt\U\st\lb{uaa}
\ea
where $\U$ is the matrix (\ref{uij}), and $\st$ is formed from $\sk$ in the same way
as $\ct$ is from $\ck$ (by discarding the wavelet part outside the same region $D$).
For the approximate energy we make a similar ansatz:
\ba
U_\f ^\a  =\ctt\Uf\st\lb{ufa}
\ea
where the matrix elements of $\Uf$ are given by (\ref{ua2}).

Let us estimate the error of the energy expression (\ref{ufa}),i.e., its difference from (\ref{uaa}).
It follows from (\ref{ct1}) and (\ref{ddd}) that
\ba
 \ct=\ck-\vdd \lb{ct2}
\ea
and, similarly, for $\st$:
\ba
\st=\sk-\vbb  .\lb{st}
\ea
Plugging (\ref{ct2}) and (\ref{st}) into (\ref{uaa}),(\ref{ufa}), we get
\ba
U^\a  &=&\ckt\U \sk-\vdt\U\sk-\ckt\U\vbb   +\vdt\U\vbb;\nn\\
U^\a_\f &=&\ckt\Uf \sk-\vdt\Uf\sk-\ckt\Uf\vbb   +\nn\\ &+&\vdt\Uf\vbb.\nn
\ea
Therefore, the error of the adaptive energy approximation has the form
\ba
\de U^\a  &\eq& U^\a  -U^\a  _\f =\de U_1+\de U_2+\de U_3+\de U_4;\qq
\de U_1=\ckt\U \sk-\ckt\Uf \sk;\nn\\
\de U_2&=&\vdt\Uf\sk-\vdt\U\sk;\qq
\de U_3=\ckt\Uf\vbb-\ckt\U\vbb;\nn\\
\de U_4&=&\vdt\U\vbb-\vdt\Uf\vbb.\lb{deu2}
\ea
One can show that all four error terms are small: 
\ba
\de U^\a  =\OO(h^{2m}).\lb{deua}
\ea
 First,
$\de U_1=\OO(h^{2m})$ because this is the error term in the non-adaptive case.
Then, let us consider the term
\ba
\de U_2=\vdt(\Uf-\U)\sk.\nn
\ea 
 One can show that it contains the non-adaptive
gradients:
\ba
\Uf\sk=\Uf\sum_j s_j^k|\phi_j^k>=\sum_{i,j}|\phi_i^k>U^f_{ij}s_j^k=
\sum_i |\phi_i^k>{\d U_f\over \d c_i^k}\nn
\ea
and the same for $\U\sk$. Therefore,
\ba
(\Uf-\U)\sk=\sum_i |\phi_i^k>\biggl({\d U_f\over \d c_i^k}-{\d U\over \d c_i^k}\biggr)=
\OO(h^{2m})\nn
\ea
because of (\ref{ggf}). Also, the wavelet 
coefficients asymptotically behave as 
\ba
d_i^{k-1}=\OO(h^{m+1/2}),\qq {\rm so}\qq 
|d^{k-1}>=\OO(h^{m})
\lb{do}
\ea 
(see \cite{Sweld1}). Therefore, 
$\de U_2=\OO(h^{3m})$ and this term can be neglected. 
Since the potential energy matrices (\ref{uij}), (\ref{ua2}) are Hermithean,
the same applies to the third term.

Eq. (\ref{do}) also determines the asymptotic behavior of the fourth term: $\de U_4=\OO(h^{2m})$. The fourth error term is formed by wavelet coefficients on the $k-1$ level,
so it is the main source of error. However, it vanishes outside $D$. On the other hand,
inside $D$ one can use the non-adaptive error estimate (\ref{uuf}). Thus the main contribution
to the energy error comes from the boundary of $D$ where neither of the above arguments applies.

An interesting partial case is when the fine region $D$ is empty. Then, all the wavelets
on the finest level $k$ are discarded, so the wavelet vector (\ref{ddt}) is zero.
Thus Eq. (\ref{ct1}) is simplified to $\ct=\ckm$,
which is equivalent to using just the scaling functions on the level $k-1$.
However, we still use the quadrature (\ref{wl}) on the level $k$, and evaluate the wave function on the grid with constant $h=2^{-k}$, which is twice denser than the grid on which
the scaling functions are defined. This is useful if we want to determine the potential
energy more precisely without raising the number of the degrees of freedom in the variational
minimization.

\subsection*{\it 5.2\quad The matrix elements for the adaptive case.}

If one substitutes (\ref{ct2}),(\ref{st}) directly into (\ref{uaa}), one obtains:
\ba
U^\a  &=&\ckmt\U \skm+\dtt\U\skm+
\ckmt\U\tbb+\nn\\&+&\dtt\U\tbb;\lb{uadt}\\
U^\a  _\f &=&\ckmt\Uf \skm+\dtt\Uf\skm+
\ckmt\Uf\tbb+\nn\\&+&\dtt\Uf\tbb\lb{ufadt}
\ea
with the elements of the exact matrix having the form
\ba
u_{ij}^{cc}\eq<\phi_i^\km|\U|\phi_j^\km>=\int \phi_i^\km(x)V(x)\phi_j^\km(x)dx;\nn\\
u_{ij}^{cd}\eq<\phi_i^\km|\U|\psi_j^\km>;\qq
u_{ij}^{dd}\eq<\psi_i^\km|\U|\psi_j^\km>.\nn
\ea

The approximate matrix elements are a bit more difficult to derive: using the backward transformation
and (\ref{ua2}), we get:
\ba
u_{ij}^{cc\,\f}&\eq&<\phi_i^\km|\Uf|\phi_j^\km>=
\sum_t V_{t}^k v_{t-2i}v_{t-2j}=
\sum_s V_{2i+s}^k v_{s}v_{s-2q}\lb{vv}
\ea
where $q=j-i$, and a new filter is defined:
\ba
v_s\eq\sum_l h_l w_{s-l}.\lb{vs}
\ea

Suppose that $V(x)=x^t$. Then, we have (\ref{kqp}) and an analog of (\ref{ufk}):
\ba
u_{0q}^{cc}=(2h)^tK_{qt};\qq
u_{0q}^{cc\f}=(2h)^tk_{qt}^f;\qq
k_{qt}^\f&\eq&\sum_l (s/2)^tv_sv_{s-2q}.\lb{kpqf}
\ea

For a general potential, we have (\ref{uk})
\ba
u_{0q}^{cc}=\sum_{p=0}^\pp{V^{(p)}(0)\over p!}(2h)^pK_{qp}+\OO(h^{2m}) .\lb{uk2}
\ea
For other matrix elements the same formula can be used, just with wavelets instead 
of scaling functions in the definition (\ref{kqp}) of  $K_{qp}$.

If we use a variant of the Taylor series (\ref{vt}):
\ba
V_{s}^k=V(sh)=\sum_{p=0}^\pp{V^{(p)}(0)\over p!}(2h)^p(s/2)^p+\OO(h^{2m})\lb{vt2}
\ea
and combine it with (\ref{kpqf}), we will obtain an analog of (\ref{uk2}):
\ba
u_{0q}^{cc\,\f}&=&\sum_{p=0}^\pp{V^{(p)}(0)\over p!}(2h)^pk_{qp}^\f+\OO(h^{2m}).\lb{ufk2}
\ea
The matrix elements containing wavelets have the same form, the only difference being that
the filters $g_i$ are used instead of $h_i$ where appropriate.

\subsection*{\it 5.3\quad The gradient error for the adaptive approximation.}

The derivatives of (\ref{uadt}),(\ref{ufadt}) w.r.t. $c_i^\km$ and $\dt_i^\km$ have the form:
\ba
{\d U^\a  \over \d c_i^\km}=\sum_j u_{ij}^{cc}s_j^\km+\sum_j u_{ij}^{cd}\bt_j^\km=
{\d U\over \d c_i^\km}-\sum_ju_{ij}^{cd}\vb_j^\km;\lb{ga1}\\
{\d U^\a  \over \d \dt_i^\km}=\sum_ju_{ij}^{dd}\bt_j^\km+\sum_ju_{ji}^{cd}s_j^\km=
{\d U\over \d \dt_i^\km}-\sum_j u_{ij}^{dd}\vb_j^\km\lb{ga2}
\ea
where $U=\ckt\U\sk$ is the energy with all wavelets kept, i.e., for the non-adaptive case. As in the non-adaptive case, the expressions for the gradient of a quadratic form
would differ just by a factor of 2.

For the approximate energy, the gradient expression is similar. The matrix elements for the exact
and approximate energy have the form (\ref{uk2}),(\ref{ufk2}), respectively. Since the filters
$K^p_q$, $k^p_q$ have finite and short length, the gradient expressions (\ref{ga1}),(\ref{ga2}) are 
nearly local. Therefore, one can consider their features depending on the region where $c_i^\km$ is located.

In the fine region $D$ (see (\ref{ddt}) and the text after it)
the $\vb_i^\km$ wavelets (\ref{ddisc}) are zero. Therefore, Eqs. (\ref{ga1}),(\ref{ga2}) reduce to their  non-adaptive counterparts. If we apply the backward wavelet transform to them, the result will be
(\ref{grad}). Then it follows from (\ref{ggf}) that the difference between the gradients of the exact and approximated
energy will behave like $\OO(h^{2m})$. Also, all the arguments of subsection 4.3. apply.

Now there is the region far from $D$ where all the wavelets are discarded.  We will also call it "coarse region".
The wavelets (\ref{ddt}): $\dt_i^\km$,$\bt_i^\km$ are zero there. 
Thus, in the coarse region (\ref{ga1}) assumes the form
\ba
{\d U^\a  \over \d c_i^\km}=u_{ij}^{cc}s_j^\km.\lb{uag}
\ea
The condition (\ref{ga2}) is not applicable in the coarse region, since the $\dt_i^\km$ wavelets are zero there, and the potential energy does not depend on them.

We will prove below that in the coarse region,
\ba
{\d U^a\over\d c_i^\km}={\d U_\f^a \over\d c_i^\km}+\OO(h^{2m+1/2}).\lb{ggfc}
\ea
In the partial case when $D=\emptyset$ discussed in the end of subsection 5.1, the above 
equation plays the role of (\ref{ggf}).

In the general case there is also the border region between the fine and coarse ones. 
The error estimates (\ref{ggf}),(\ref{ggfc}) were derived under the assumptions that either
the $\dt_i^k$ or $\vd_i^k$ wavelets are zero, correspondingly. Since in the border region
neither of these assumptions is guaranteed, the estimates (\ref{ggf}),(\ref{ggfc}) may no
longer hold there (and the numerical tests imply that they do not hold indeed). 
Thus in the border region, one has just (\ref{ga1}),(\ref{ga2}), with all terms being nonzero. From the right equations
of (\ref{ggf}),(\ref{ggfc}) one can conclude that the error at the border is bounded by
 the magnitude of $\vb_j^\km$ coefficients which behave according to (\ref{do}). The 
other, non-adaptive-like part of error behaves like $\OO(h^{2m})$, because of (\ref{ggf}).
 Therefore, 
\ba
{\d U\over\d c_i^\km}={\d U_\f \over\d c_i^\km}+\OO(h^{m+1/2});\qq
{\d U\over\d \dt_i^\km}={\d U_\f \over\d \dt_i^\km}+\OO(h^{m+1/2})\nn
\ea
in the border region, so it is the main source of error.
The small energy error (\ref{deua}) is consistent with the above because
  the rapidly oscillating part of the gradient at the boundary 
region has small smooth component and gets drastically diminished when multiplied by a smooth
function, according to (\ref{ugrad}).

In the remaining part of this subsection we will prove (\ref{ggfc}) in the coarse region.
The proof is completely analogous to that of (\ref{ggf}), so we will not go into much detail. First, let $V(x)=x^t$ and $\P(x)=x^l$. Then, it follows from (\ref{uu}) that
\ba
{\d U^a\over\d c_0^\km}=(2h)^{t+l+1/2}M_{l+t}.\nn
\ea
Let us prove that the same is true for $U_\f^a$. 
Substituting the potential energy matrix from (\ref{kpqf}) and (\ref{cjk2}) for the level $k-1$ into (\ref{uag}) we obtain the following analog of (\ref{af}):
\ba
{\d U_\f^a\over\d c_0^\km}=(2h)^{t+l+1/2}a_{lt}^\f;\qq
a^\f_{p,u}\eq\sum_{t=0}^p C_p^tM_t\sum_q k_{qu}^\f q^{p-t}.\nn
\ea

Similarly to (\ref{uu}),one can show that
\ba
a_{lt}^\f=M_{l+t};\qq
{\d U_\f^a\over\d c_0^k}={\d U^a\over\d c_0^k}=(2h)^{l+t+1/2}M_{l+t};\qq l+t<2m.\lb{affm}
\ea
For the values $l<m$, Eq. (\ref{affm}) can be proved along the lines of Appendix B. 
Since we were unable to find an analytical proof of (\ref{am}) for 
the remaining values of $l$, 
we checked it numerically with Mathematica for the Daubechies wavelets of the orders from 6 to 18. We checked both the extremal phase and least asymmetric families.
Same as with (\ref{amm}), the relative error of (\ref{affm}) can be made arbitrarily small by increasing the precision
of wavelet filters. We went down to $10^{-40}$.

Combining (\ref{affm}) with the Taylor expansions (\ref{vt2}) and (\ref{pt}) for the
level $k-1$, we get an analogue of (\ref{gaa}):
\ba
{\d U_\f^a \over\d c_0^\km}&=&
\sum_{s=0}^\pp {(2h)^{s+1/2}\over s!}
\sum_{p=0}^sC^p_s\P^{(p)}(0)V^{(s-p)}(0)a^\f_{p,s-p}+\OO(h^{2m+1/2})=\nn\\
&=&\sum_{s=0}^\pp {(2h)^{s+1/2}\over s!}G^{(s)}(0)M_s+\OO(h^{2m+1/2})=
g_0^{k-1}+\OO(h^{2m+1/2})=\nn\\
&=&{\d U^a \over\d c_0^\km}+\OO(h^{2m+1/2}).\lb{fgaa2}
\ea
Eq. (\ref{ggfc}) is thus proved.

\section*{6. \quad Reducing the computational cost in the coarse region.}

\subsection*{\it 6.1 \quad The adaptive quadrature for the product of a scaling function
and a smooth function.}

The approximation derived above is not numerically efficient in the coarse region. Namely,
the approximate energy matrix elements are defined by (\ref{uvec}) there. However,
in the non-adaptive approximation for the level $k-1$, they would be given by (\ref{vv}) ,
which requires roughly twice smaller number of calculations because the filter $\w_l$ is
shorter than $v_l$ The asymptotic behavior of the energy error of these two approximations is the same: $\OO(h^{2m})$. Therefore, it would be desirable to use (\ref{vv}) in the coarse region.

This situation is analogous to applying the quadrature (\ref{wl}) on the level $k$
to a two-scale function $G(x)$ that has "fine region" where one should use the quadrature
(\ref{wl}) on the level $k$ and "coarse region" where one should use it on the level $k-1$.

Thus we have an adaptive quadrature. Recall that the adaptive approximation considered
in the previous Sections consists of setting the wavelet coefficients in the coarse region
to zero. In addition to that, the adaptive quadrature is defined on a coarser grid
in the coarse region.

The question is - what quadrature should we use at the boundary of the coarse and fine regions?
Our recipe is to use the fine quadrature, along with backward 
transformation, with the filter (\ref{vs}):
\ba
\int G(x) \phi_r^{k-1}(x)dx&=&\sqrt{h}\sum_lv_lG_{l+2r}^k+\OO(h^{M+3/2})=
\sqrt{h}\sum_s v_{s-2r} G_s^k+\nn\\
&+&\OO(h^{M+3/2})\nn
\ea 
and go to the coarse quadrature only at some distance from the boundary.
 
 An alternative (and maybe more rigorous) recipe \cite{John} is to use
a non-uniform quadrature at the boundary. However,
\begin{itemize}
\item The non-uniform quadrature is difficult to program since the points in space
near the boundary should be dealt with individually.
\item  Our error estimates for the potential energy quadrature hold only for the uniform case and it is not trivial to extend them onto the case of non-uniform quadrature of \cite{John}.
\end{itemize}

\subsection*{\it 6.2 \quad The energy gradient in the coarse region.}

At present, the approximate energy gradient in the coarse region is given by
\ba
{\d U^\a  \over \d c_i^\km}=\sum_j u_{ij}^{cc\,\f}s_j^\km\nn
\ea
where the matrix elements are given by (\ref{vv}). In general , the gradient is given 
by (\ref{ga1}),(\ref{ga2}).

The energy gradient on the $k-1$ level in the non-adaptive scheme is given by
(\ref{grad}) with the matrix elements (\ref{ua2}) (for the level $k-1$):
\ba
{\d U_\f \over\d c_i^\km}=\sum_{j}s_j^\km U_{ij}^{\f \,k-1}.\lb{guf}
\ea

One could define a "quasigradient"
vector $v_i^\km$ with the following components:
\ba
v_i^\km=\left[\matrix{{\d U^\a  _\f \over \d c_i^\km}=\sum_ju_{ij}^{cc\,\f}s_j^\km+\sum_ju_{ij}^{cd\,\f}\bt_j^\km ,i\in D'     \cr
       {\d U_\f \over\d c_i^\km}=  \sum_{j}s_j^\km U_{ij}^{\f \,k-1},i\out D'  }\right.
\lb{vi}
\ea
where the set $D'$ is $D$ plus the points that are no more than a distance $a$ (empirically, $a=3mh$ is enough) away from $D$. The word "quasigradient" means that the vector (\ref{vi})
is assembled from gradients (\ref{ga1}), (\ref{guf}), but is not necessarily equal
to the gradient of any function at all.

We can use now use this vector instead of the gradient in the iterative minimization
algorithms.

The quasigradient vector (\ref{vi}) is a good approximation of the adaptive gradient (\ref{ga1}).
For the points $i\in D'$ their components coincide. For the points $i\out D'$,
they coincide if $V(x)$ and $\P(x)$ are polynomials with the sum of their degrees
smaller than $2m$, because in that case both the quasigradient and the adaptive gradient coincide
with the exact gradient (see (\ref{uu}),(\ref{affm})).

For the general potentials and wave functions, the gradient (\ref{ga1}) assumes the form (\ref{fgaa2}). On the other hand,
the quasigradient will have the form (\ref{gaa}) for the level $k-1$, which is identical to 
(\ref{fgaa2}).
 Therefore, the quasigradient components
are asymptotically close to those of the adaptive gradient (\ref{ga1}) and 
its approximate version:
\ba
{\d U_\f ^\a  \over\d c_i^k}-v_i^k=\OO(h^{2m+1/2});\qq 
{\d U^\a  \over\d c_i^k}-v_i^k=\OO(h^{2m+1/2}).\lb{grad2}
\ea

Now let us consider the energy.  Eq. (\ref{ufa}) can be rewritten in the form
similar to (\ref{ugrad}):
\ba
U_{f}^\a  =\sum_i c^\km_i{\d U_\f \over \d c^\km_i} +\sum_i \dt^\km_i {\d U_\f \over \d \dt^\km_i}.\nn
\ea
One can define a likewise energy expression based on (\ref{vi})
\ba
U_{e}^\a  =\sum_i c^\km_i v^\km_i+\sum_i \dt^\km_i {\d U_\f \over \d \dt^\km_i}\lb{ue}
\ea
where "e" stands fort for "efficient". Then it follows from (\ref{grad2}) that
\ba
U_{e}^\a  =U_\f ^\a  +\OO(h^{2m});\qq U_{e}^\a  =U^\a  +\OO(h^{2m}).\nn
\ea
Thus, the energy expression (\ref{ue}) is also a good approximation of (\ref{ufa}), (\ref{uaa}). 

Note that the gradient of (\ref{ue}) does not coincide with the quasigradient (\ref{vi}). They are different only at the boundary of $D'$: the former behaves badly there, while the latter is smooth. This means, in particular, that simply minimizing (\ref{ue})
would result in a wavefunction that is not smooth at the boundary of $D'$ , but oscillates there
rapidly.
 
 The quasigradient can be seen as a projection of the
gradient of (\ref{ue}) onto the space of functions that are smooth at the boundary of $D'$.
One can then assume that we minimize the energy (\ref{ue}) in the space of such functions.
This remedies the above problem of singularity at the boundary.

Alternatively, one can consider the quasigradient as an approximation of the gradient
(\ref{ga1}) and (\ref{ue}) as an approximation of (\ref{ufa}). We have shown above that
the error of such approximation is asymptotically small for smooth wavefunctions and potentials.

\section*{7. \quad Charge density and products of functions.}

In the density functional theory one needs to express the charge density 
\ba
\rho(x)=\biggl(\P_I(x)\biggr)^2\nn
\ea
as a linear combination of scaling functions. In the Hartree-Fock method, one uses 
products of functions of the form (\ref{pi1}):
\ba
\F_I(x)\P_I(x)=\sum_{i,j} c_i^ks_j^k\phi_i^k(x)\phi_j^k(x)\lb{fp}
\ea
and it would also be useful
to have that product expanded in terms of scaling functions or wavelets, not in
terms of scaling function products:
\ba
\F_I(x)\P_I(x)=\sum_jf_j^k\phi_j^k(x)+\de F(x)\nn
\ea
where $\de F(x)$ is small. In general, one can expand the product in
scaling functions of the order different from $m$ (that for $\F_I(x),\P_I(x)$), or
even from some other wavelet family.

The obvious way to obtain the coefficients $f_j^k$ is to expand the left part
in scaling functions according to (\ref{fw}). The coefficients can then be obtained
by convolution with the matrix of triple products, similarly to (\ref{ui}).
However, then we would have the same problem as with (\ref{ui}): too many flops
per grid point, $~N^2$ in one dimension and $~N^4$ in three dimensions, where $N$ is the dimension of the triple product matrix.

Fortunately, we can suggest an
alternative way to get the density:

\subsection*{\it 7.1. \quad Using the average grid values of $\F(x)$ and $\P(x)$.}

Let us define the following coefficients:
\ba
F_i^k\eq \bar\F_i^k\bar\P_i^k=\F(ih)\P(ih)+\OO(h^{2m})\nn
\ea
where the average values are defined in (\ref{vpq}). 

Then one can approximate the product (\ref{fp}) with the function
\ba
F_\de(x)=h\sum F_i^k\de(x-ih).\lb{fde}
\ea
One can easily show that this function reproduces the multipole moments of the product 
(\ref{fp}):
\ba
\int F_\de(x) x^t dx=h\sum_j \F(jh)\P(jh) j^t= \int \F_I(x)x^t\P_I(x)dx+\OO(h^{2m})\lb{fxt}
\ea
for all integer $t\ge 0$. The second equality is a partial case of (\ref{uuf}).

What is more, if $\P(x)$ is a polynomial of the degree $l$ and $l+t<2m$, then
(\ref{fxt}) is satisfied {\it exactly}. This is a consequence of (\ref{uu}).

However, the function (\ref{fde}) is not smooth, in contrast to
the product (\ref{fp}) that it should approximate. To remedy this, one can go from (\ref{fde})
to
\ba
F_I(x)=\sum F_i^k\phi^{Ik}_i(x);\qq
\phi^{Ik}_i(x)\eq \phi^I(x/h-i)\nn
\ea
where $\phi^I(y)$ is the interpolating scaling function of the order $L$. Its first
$L$ moments are the same as those of a delta function \cite{beyl2},\cite{tian}. Therefore,
$F_I(x)$ satisfies (\ref{fxt}) too (for $t<L$).

On the other hand, if $L\geq m$, then
\ba
F_I(x)=\F(x)\P(x)+\OO(h^L)=\F_I(x)\P_I(x)+\OO(h^m)\nn
\ea
so the approximated density is close to (\ref{fp}) at any point.

This approximation is much faster than that the triple product calculation. However, the moment
conservation (\ref{fxt}) is not exact. To make at least the total charge 
of a single electron equal to one in this scheme, one has to scale the
coefficients $F_i^k$ accordingly. One can also make a forward wavelet transformation 
to the scaling function coefficients on some coarser level and scale only them; this is
enough to get the correct total charge.

\section*{8. \quad Application to the harmonic oscillator.}

\subsection*{\it 8.1. \quad The non-adaptive case.}

In this Section we will test the above approximations by finding the ground state of a unit 
mass and frequency harmonic oscillator in one dimension. We use nonperiodic boundary conditions 
where the wavefunction was set to zero outside the interval [-16:16]. 
The variational ground state is obtained from Eq. (\ref{min}), with the definitions (\ref{RR}),(\ref{hpsi}) and $V(x)=x^2/2$. Since for the oscillator potential, (\ref{vint})
is exact, we can use the triple product method (\ref{ui}) without invoking additional errors.
  
Our approximation for the potential energy has the form (\ref{ufd}):
\ba
U_\f(\{c_i\}) = h\sum_{s}V^k_s\biggl(\bar\P^k_s\biggr)^2;\qq 
\bar\P_q^{k}={1\over\sqrt{h}}\sum_t W_{q-t}c^{k}_t.\nn
\ea 
Accordingly one can define the approximate RR functional
\ba  
R_\f(\{c_i^k\})&=& {T(\{c_i^k\})+U_\f(\{c_i^k\})\over \ckt c^k>};\qq
E_0^\f=\min R_\f(\{c_i^k\});\lb{Rf}\\ |c^k_\f>&=&\arg\min R_\f(\{c_i^k\}).\nn
\ea

We performed minimization of (\ref{RR}) in the exact and approximate
cases for least asymmetric Dau\-be\-chies - $2m$ with $3\leq m\leq 8$. We used the steepest descent method with diagonal preconditioning and gradient feedback. The results for Daubechies-8 and Daubechies-16 are shown on Figs 2-5. On the $x$ axis we have the inverse grid interval:
$h^{-1}=2^k$. 

In Figs 2,3, on the $y$ axis we have the deviation of the variational ground state energy $E_0$ (\ref{min}) from the exact result (which is $0.5$ for the unit oscillator) and the deviation of the approximate ground state energy $E_0^\f$ (\ref{Rf}) from $0.5$.
 
Also shown is the difference of the variational and approximate ground state energies
(\ref{Rf}) and (\ref{min}):
\ba
\de E_{app} \eq E_0^\f-E_0\nn
\ea
 which we call the approximation error. The graphs are on a double logarithmic scale,
and they have a distinct linear part. The slope of the linear part is consistent with
(\ref{eex}). For the lower orders, the approximation error 
is one or two orders of magnitude smaller than (\ref{eex}). Its slope in the linear region
is equal to $2m$, suggesting that
\ba
\de E_{app}=\OO(h^{2m}).\lb{deeint}
\ea
This behavior is similar to (\ref{uuf}). However, note that here we compare the energies
for different (although close) quantum states $|c_0^k>$ and $|c_\f^k>$ defined 
in Eqs. (\ref{min}) and (\ref{Rf}), while in the
Section 3 we computed the exact and approximate energy for the same state.

The approximation
error grows with $m$ faster than the variational error, so that for Daubechies-16 they
become of the same magnitude, and have the same slope $2m-2$. 

In Figs 4,5, on the $y$ axis we have the following quantities: 
\ba
\de c^k=\sqrt{\sum_i (c_{i\, 0}^k-c_{i\,\rm g}^k)^2};\qq
\de c_{app}^k=\sqrt{\sum_i (c_{i\, 0}^k-c_{i\,\rm f}^k)^2}.\lb{dec}
\ea
We will call the first quantity the variational error and the second one 
the approximation error. The coefficients
\ba
c_{i\,\rm g}^k\eq C\int \exp(-x^2/2)\phi_i^k(x)dx \nn
\ea
correspond to the scaling function expansion of the exact ground state of the
unit oscillator in the space $\VV_k$ (\ref{vwk}). The factor $C$ is chosen such that
$\sum_i (c_{i\,\rm g}^k)^2=1$.

We could not find the evaluation of the quantities (\ref{dec}) neither in \cite{Strang}
nor in the previous papers describing the application of wavelets
to the Schroedinger equation \cite{John0},\cite{John}. The quantity of interest
in \cite{Strang} was something else: the norm  of the total difference between the exact ground 
state of (\ref{hpsi}) (in this case, a Gaussian) and $\sum_i c_i^k\phi_i^k(x)$:
\ba
\| \pi^{-1/4}\exp(-x^2/2)- \sum_i c_i^k\phi_i^k(x)\|=\OO(h^m).\lb{ertot}
\ea
The above quantity includes the fine wavelet part (\ref{defo}) of the Gaussian that also 
behaves as $\OO(h^m)$. 
 
For lower values of $m$,
the slope of the graphs of (\ref{dec}) in the linear region is $2m-2$ and $2m$
correspondingly. The approximation error in the linear region is again
one or two orders of magnitude smaller than the variational error. This
suggests the asymptotic behavior
\ba
\de c^k=\OO(h^{2m-2});\qq \de c_{app}^k=\OO(h^{2m}).\lb{dech}
\ea
It follows from the first estimate (\ref{dech}) that the wave function error (\ref{ertot})
is dominated by the fine wavelet part.
The second estimate of (\ref{dech}) means that the minima of the exact and approximate RR functionals are
very close both to each other and to the projection of the exact solution onto the space of the scaling functions 
at resolution level $k$.  
This explains in part why the energy error (\ref{deeint}) behaves similarly to (\ref{uuf}).

For higher values of $m$ the error reaches  the machine precision range before
 the slopes reach their asymptotic values. Still, the approximation error decays
 faster than the variational one, and the slopes differ by $2$, approximately.

For the lower values of $m$ there is no difference between the behavior of the least asymmetric and extremal phase Daubechies scaling functions. 
However, for $m>10$ the approximation error for the extremal phase Daubechies approaches
the variational error. The convergence of the steepest descent iterations becomes 
very bad. For the symmetric family nothing of that happens.
Thus for the bigger values of $m$ our method is not applicable for the extremal
phase family. This is the reason why we prefer the least asymmetric family in general.

In a practical electronic structure calculation the exact minimization is too costly numerically, but one still needs a criterion of accuracy of our approximation. We can use for that purpose the gradient of the exact RR functional (\ref{RR}) at the minimum of the approximate one (\ref{Rf}).

\subsection*{\it 8.2. \quad The adaptive case.}

We will impose adaptivity in the following way: the minimum of the RR functional
(\ref{RR}) will be sought in the class of coefficients $\{c_i^{k-1},\dt_i^{k-1}\}$ such that
the wavelet coefficients are zero outside the
interval $[-\kappa,\kappa]$: $\dt_i^{k-1}=0$ for $|2^{1-k}i|\geq\kappa$. 
We choose $\kappa=1.5$ as an illustration. 
The resulting energy and wavefunction are,
\ba
E_0=\min R(\{c_i^{k-1},\dt_i^{k-1}\})=\min R(\{\tc_i^k\});\qq 
|\tc^k_0>=\arg\min R(\{\tc_i^k\})\nn
\ea
where the coefficients ${\tc_i^k}$ are the backward transformation of $\{c_i^{k-1},\dt_i^{k-1}\}$.

Their approximate counterparts are
\ba
E_\f=\min R_\f(\{c_i^{k-1},\dt_i^{k-1}\})=\min R_\f(\{\tc_i^k\});\qq 
|\tc^k_\f>=\arg\min R_\f(\{\tc_i^k\})\nn
\ea
where in the approximate RR functional the potential energy part is treated
according to the Sections 5-6.

The graphs of the energy and wavefunction error for the (least asymmetric) Daubechies-8 and Daubechies-16 are shown on Figs 6-9. The main differences from  the non-adaptive case are:
\begin{itemize}
\item The slope of the variational wavefunction error is now approximately $m+1/2$.
The asymptotical slope of the approximation error exceeds that of the variational error 
by 2, roughly. Both wavefunction errors are localized at the boundary.
\item
The approximation error both for the energy and wave function for small wavelet orders is an order of magnitude smaller than the variational error, now for the bigger values of $h$ too.
\item
The behavior of the approximation error is improved for the small $k$, compared to
the non-adaptive case.
\end{itemize}

\section*{9. \quad Conclusion.}

In the present work we propose a quadrature for the evaluation of the potential energy 
functional when the wave function is a wavelet approximation of some smooth function. We used the Daubechies family but the results can be extended onto others.
With our algorithm,
the potential energy can be calculated using only one-dimensional convolutions and filters,
in contrast to the existing methods. The resulting potential energy differs only
insignificantly from the exact value. The algorithm is extended onto the case of varying
spatial resolution (adaptivity). As a numerical test we calculated the ground state energy
and wave function of the harmonic oscillator in 1 dimension for the least asymmetric  and extremal phase Daubechies wavelets with
orders from 6 to 16. We performed the minimization of the RR functional with the potential energy calculated by our method and compared the resulting energy and wavefunction with those obtained from the fully variational minimization. In the case of the
least asymmetric Daubechies family the approximate energy and wavefunction are close to the variational values. However, for the extremal phase family our method is reliable only
for wavelet orders less than $10$.

Our method also allows a fast calculation of a charge density for a wavefunction 
expressed in the wavelet basis. As a byproduct we derived a filter for reconstruction
of the grid values of a function from its Daubechies-$2m$ scaling function expansion. This
reconstruction is exact for polynomials up to order $2m$ and the length of the filter is just $2m$.

The method can readily be generalized to 3 dimensions and it is already being
used for 3-dimensional electronic structure calculations in the framework of the BIGDFT project \cite{bigdft} which is a subject of future publication.

\section*{10. \quad Acknowledgments.}

The authors are grateful to R. Schneider for useful discussions.

\section*{ Appendix A.\quad Reconstruction of grid values of wave function.}

The aim of this subsection is the reconstruction of the value of a smooth function at the grid points from the scaling function expansion coefficients.
For this purpose we need to find the finite (and shortest possible) length filters $\W_l$ 
such that
\ba
\bar\P_r^k\eq\hh\sum_s \W_{r-s}c^k_s=
\hh\sum_s \W_qc^k_{r-q}=\P(hr)+\OO(h^{2m})\lb{aac}
\ea
where $c_i^k$ are given by (\ref{cik}) and $h=2^{-k}$. 

One can start by finding filters that satisfy (\ref{aac}) exactly for $
\P(x)=x^l$, $l<2m$ and $k=0$:
\ba
\sum_s W_{r-s}c_s[x^l]=r^l\lb{wrl0}
\ea
where the coefficients $c_s[x^l]$ are a generalization of the moments (\ref{Ms}):
\ba
c_s[x^l]\eq \int x^l \phi_s(x)dx.\nn
\ea

Eq. (\ref{wrl0}) means that, though with the usual interpolation (\ref{pq})
using  Daubechies-$2m$ scaling functions one can exactly reproduce polynomials
of degree not more than $m-1$, the desired filter $W_l$ should allow us
to reproduce polynomials of degrees up to $2m-1$. 
Thus for such high-order polynomials, (\ref{aac}) would be satisfied exactly,
so the average value $\bar\P_r^k$ would coincide with value $\P_r(rh)$ of the polynomial,
but {\it not} with the value of the scaling function expansion (\ref{pq}) 
\ba
\P_I(rh)=\sum_i \phi_i^k(rh)\int \phi_i^k(y)\P(y) dy  .\nn
\ea
It is enough to prove (\ref{wrl0}) at the origin only:
\ba
\delta_l=\sum_s W_{-s}c_s[x^l]\lb{dew}
\ea
since then it would follow that
\ba
\sum_s W_{r-s}c_s[x^l]&=&\sum_q W_{-q} c_{q+r}[x^l]=
\sum_q W_{-q} c_{q}[(x+r)^l]=\nn\\
&=&\sum_q W_{-q}\sum_{u=0}^lC_l^ur^{l-u}c_q[x^u]=
\sum_{u=0}^lC_l^ur^{l-u}\de_u=r^l\lb{wrl}
\ea
where we changed the indices: $q=s-r$.
To prove (\ref{dew}), we will use the following auxiliary formula:
\ba
c_s[x^l]&\eq& \int x^l \phi_s(x)dx=\int x^l \phi(x-s) dx=\int (y+s)^l\phi(y)dy=\nn\\
&=&\int \sum_{u=0}^lC_l^uy^{l-u}s^u\phi(y)dy=\sum_{u=0}^lC_l^us^uM_{l-u}\lb{csm}
\ea
where $C_l^u$ are the binomial coefficients, and $M_s$ are the scaling
function moments (\ref{Ms}).

Substituting (\ref{csm}) into the right part of (\ref{dew}),we get:
\ba
\sum_s W_{-s}c_s[x^l]&=&\sum_s W_{-s}\sum_{u=0}^lC_l^us^uM_{l-u}=
\sum_{u=0}^lC_l^uM_{l-u}\ga_u(-1)^u=\de_l;\lb{gade}\\
\ga_u\eq \sum_s W_s s^u.\lb{gau}
\ea
The filter $W_s$ can thus be found by solving the systems
of linear equations (\ref{gade}),(\ref{gau}) for $\ga_u$ and then for $W_s$.

Now we are ready to prove (\ref{aac}).
It is enough to do it only at the origin:
\ba
\bar\P_0^k=\hh\sum_s \W_{-s}c^k_s=\P(0)+\OO(h^{2m}).\lb{psi0}
\ea

To prove (\ref{psi0}) one should use a version of (\ref{dew}) at the level $k$:
\ba
\delta_l={1\over\sqrt{h}}\sum_s W_{-s}c_s^k[x^l]\lb{dehw}
\ea
that can be proved in the same way as (\ref{dew}) itself.

Then it remains to expand $\P(x)$ in the Taylor series at $0$:
\ba
\P(x)=\sum_{p=0}^{2m-1}{\P^{(p)}(0)\over p!}x^p+\OO(h^{2m})\nn
\ea
and plug it into (\ref{psi0}):
\ba
&&{1\over\sqrt{h}}\sum_s W_{-s}c_s^k[\Psi(x)]=
{1\over\sqrt{h}}\sum_s W_{-s}c_s^k\biggl[
\sum_{p=0}^{2m-1}{\P^{(p)}(0)\over p!}x^p+\OO(h^{2m})
\biggr]=\nn\\
&=& {1\over\sqrt{h}}\sum_s W_{-s}\sum_{p=0}^{2m-1}{\P^{(p)}(0)\over p!}c_s^k[x^p]+\OO(h^{2m})=
\P(0)+\OO(h^{2m}),\nn
\ea
the last equality following from (\ref{dehw}). Eq. (\ref{psi0}) is thus proved.

Suppose we work with Daubechies-$2m$ scaling functions. Then,  Eqs. (\ref{gade}) are satisfied if
\ba
\ga_u= M_{u},\qq u=0,\ldots,\pp \lb{lem3}
\ea
because of the equality ((B2) from \cite{John}):
\ba
\sum_{s=0}^pC^s_pM_{p-s}M_s(-1)^s=M_p^I=\de_p;\qq p=0,\ldots,2m-1\lb{mmm}
\ea
where $M_p^I$ are the moments of the lazy-$m$ scaling function.
The last equality is a property of the interpolating scaling functions \cite{beyl2},\cite{tian}. Eq. (\ref{mmm}) is satisfied both for the
extremal phase and least asymmetric Daubechies wavelets.

The shortest filter with the  moments (\ref{lem3}) is (\ref{wl}). Thus, 
(\ref{uuf}) reduces to (\ref{ufd}). Note however that (\ref{mmm}) is no longer 
satisfied for $p\geq 2m$ because the higher moments of an interpolating scaling function
are not zero. Therefore one cannot construct a filter of degree $p$ higher than $2m-1$
that would also satisfy (\ref{gade}) for the powers of $x$ up to the $p$-th. Thus one cannot improve (\ref{ufd}) so that its error scale as $\OO(h^s)$ with $s>2m$.

The reconstruction scheme presented above can be generalized to find the values
of $\P(x)$ at arbitrary points and also to find its derivatives of arbitrary order.

\section*{ Appendix B.\quad The analytical derivation of the $A_{lt}$, $A_{lt}^\f$ coefficients}

In this Appendix we will present the proofs of Eqs. (\ref{am}),(\ref{afm}).

At first let us prove (\ref{am}). Suppose, as in subsection 4.3, that $V(x)=x^t$ and
$\P(x)=x^l$. Then, combining the left equalities of (\ref{grad}) and (\ref{kqp}), we get,
for $l<m$,
\ba
{\d U\over \d c_0^k}=\sum_j s_j^k U_{0j}=\sum_j s_j^k\int \phi^k(x) x^t\phi_j^k(x) dx
=\int \phi^k(x) x^{t+l} dx= h^{t+l+1/2}M_{t+l}\lb{gm}
\ea
where we used the fact that $\sum_j s_j^k[x^l]\phi_j^k(x)=x^l$ for $l<m$. From (\ref{gm})
and (\ref{a}), we thus get the first equality of (\ref{am}).
On the other hand, if $t=0$ then (\ref{kqp}) reduces to $U_{0j}=\de_j$ and thus (\ref{grad})
has the form
\ba
{\d U\over \d c_0^k}=s_0^k[x^l]=h^{l+1/2}M_l\lb{g0m}
\ea
where the last equality is a partial case of (\ref{sjk}). Combining (\ref{g0m}) and (\ref{a})
we get the second equality of (\ref{am})

Now let us turn to Eq. (\ref{afm}). Plugging (\ref{ufk}):
\ba
U_{0q}^\f=h^t\sum_r r^t w_r w_{r-q}\nn
\ea
into the second equality of (\ref{grad}), we get:
\ba
{\d U_\f\over \d c_0^k}&=&\sum_j s_j^k U_{0j}^\f=h^t\sum_r r^t w_r\sum_q w_{r-q} s_q^k[x^l]=
h^t\sum_r r^t w_r \sqrt{h} (hr)^l=\nn\\ 
&=&h^{t+l+1/2}\sum_r w_r r^{t+l}=h^{t+l+1/2}M_{t+l}\lb{gfm}
\ea
where we have used (\ref{wrl}) and (\ref{lem3}). Now, Eq. (\ref{afm}) follows
from (\ref{af}) and (\ref{gfm}).

\end{document}